\documentclass[aps,prc,onecolumn,showpacs,amsmath,amssymb,nofootinbib,11pt]{revtex4-2}
\usepackage{graphicx}
\usepackage{color}
\usepackage{times}
\usepackage{inputenc}
\usepackage{bm}
\usepackage{ulem}
\usepackage{multirow}
\usepackage{float}
\usepackage{url}
\usepackage{natbib}
\usepackage{mathrsfs}
\usepackage{physics}
\usepackage{comment}
\usepackage[table,xcdraw]{xcolor}
\usepackage{booktabs,makecell}
\usepackage{comment}
\usepackage[colorlinks=true,citecolor=blue,urlcolor=blue,linkcolor=blue]{hyperref}
\begin{document}
\title{Anomalous Behavior of Giant Monopole Resonance Energy with Nuclear Matter Incompressibility in the framework of  Relativistic Mean Field Formalism and Coherent Density Fluctuation Model}

\author{Jeet Amrit Pattnaik$^{*1}$, R. N. Panda$^{**1}$, M. Bhuyan$^{***2}$, S. K. Patra$^{****1}$}
\address{$^1$Department of Physics, Siksha $'O'$ Anusandhan, Deemed to be University, Bhubaneswar 751030, India,
$^2$Institute of Physics, Sachivalaya Marg, Bhubaneswar 751005, India }
\email{*jeetamritboudh@gmail.com}
\email{**rabinarayanpanda@soa.ac.in}
\email{***mrutunjaya.b@iopb.res.in}
\email{****sureshpatra@soa.ac.in}





\begin{abstract}
The finite nucleus incompressibility $K^A$ is evaluated using the coherent density fluctuation model with the extended relativistic mean field density. The relativistic energy density functional for nuclear matter is replaced by the local density approximation for finite nuclei. The equation is used to calculate the finite nuclear incompressibility, which is further utilized to evaluate the isoscalar giant monopole excitation (ISGMR) energy $E_M$. This excitation energy is compared with other theoretical calculations and experimental data, wherever available. The results are comparable to the data. In contrast to the general understanding, the $E_M$ of finite nucleus  is found to be maximum for the lowest nuclear matter incompressibility $K_{\infty}$, whereas it is minimum for the maximum $K_{\infty}$.  These reverse results may be due to the self- and cross-interactions of the vector mesons in the nuclear potential.

\end{abstract}
\maketitle
%
The nuclear matter incompressibility at saturation $K_{\infty}$ is a crucial quantity for both the equation of state (EoS) and finite nuclei. This quantity determines the softness or stiffness of the EoS \cite{Arumugam2004}. In the study of neutron star (NS), the mass M and radius R of  NS are highly influenced by $K_{\infty}$, for higher $K_{\infty}$ the EoS is more stiff and for lower $K_{\infty}$, it is softer \cite{kuma18,kuma17}. In case of softer EoS, we get a lower mass of the neutron star, while solving the Tolman-Oppenheimer-Volkoff (TOV) equation. In the heavy ion collision (HIC) experiment, the analysis of flow data rests on the nature of the EoS, which ultimately depends on the incompressibility of the nuclear matter medium \cite{dani2002}. 

The excitation energy $E_M$ of the isoscalar giant monopole resonances (ISGMR) is a precisely measurable quantity for finite nucleus \cite{blaizot80}. The value of $E_M$ constrains the incompressibility of finite nucleus $K^A$, whose extrapolation measures $K_{\infty}$. The ISGMR are resonances with high-frequency and small-amplitude collective modes of oscillations. The study of ISGMR is a major topic of research in nuclear physics, because there is no direct way of measuring the incompressibility of a nuclear medium.  The closest experimental way to determine the $K_{\infty}$ is in an indirect way by knowing the iso-scalar giant monopole excitation energy. Since ISGMR is the mode of excitation by breathing mode, using the relation of ISGMR with the $K_{\infty}$, one can evaluate the incompressibility. One of the limitations of the method is the connection of the dynamical quantity of ISGMR with the static properties of the incompressibility of nuclear matter $K_{\infty}$. 

There are several theoretical ways to estimate the nuclear matter incompressibility. However, experimentally, one has to measure $E_M$ and then using this value of $E_M$, one can estimate $K_{\infty}$. 
A random phase approximation (RPA) in a relativistic mean-field approximation (RMF) estimates that the range of incompressibility at saturation is $K_{\infty}=270\pm{10}$ MeV \cite{Lala1997,Vret2002}. In contrast, the non-relativistic Hartree-Fock (HF) method with RPA predicts its $K_{\infty}$ to be $ 210- 220$~MeV. The recent extended relativistic mean field models \cite{kuma17,kuma18} predict the $K_{\infty}=220\pm 30$ MeV. The experimental value of the monopole resonance excitation energy is $K_{\infty}=240\pm{20}$ MeV~\cite{Colo2006, Colo2008,Colo2018,Patt2022}.
In the present work, we use the extended relativistic mean field (E-RMF) formalism along with the coherent density fluctuation model (CDFM) to evaluate $K_{\infty}$, and then we will do the back-calculation to get the ISGMR value. 

The theoretical calculation is mainly divided into two parts (i) The extended relativistic mean field formalism and the translation of the energy density functional from momentum space to the coordinate space in the local density approximation (LDA). From this relativistic energy density functional, we get the 
nuclear matter incompressibility using the standard definition  
\begin{eqnarray}
K_{\infty}&=&9\rho^2\frac{\partial^2}{\partial \rho^2} \bigg(\frac{\cal E}{\rho}\bigg)\Big|_{\rho=\rho_0} \label{knm}.
\end{eqnarray}
Where the notations are their usual meaning and can be found in Ref. \cite{kuma18,kuma17}.
(ii) The nuclear matter incompressibility is then used to evaluate the finite nuclear incompressibility using the relation\cite{pattnaik24}
\begin{eqnarray}
K^{A}= \int_0^{\infty} dx\, \vert F(x) \vert^2\, K_{\infty} (\rho (x)).
\label{k0}
\end{eqnarray}
To get the incompressibility expression for finite nuclei $K^A$, we express the  ${\cal{E}}_{nucl.}(k)$ \cite{yadav25} in the coordinate space due to the finite size of the nucleus. In other words, the energy density is reconstructed using the local density approximation (LDA) to ${\cal{E}}_{nucl.}(x)$. To convert the energy density from momentum space to the coordinate space, the nuclear matter is assumed to made up of tiny spherical pieces in the local density function called  {\it Flucton} is defined as $\rho_0(x) = 3 A/4\pi x^3$. 
The density of the considered finite nuclei are obtained from the E-RMF formalism and the weight function $|F(x)|^2$ is defined as \cite{pattnaik24}: 
\begin{equation}
|F(x)|^2 = - \left (\frac{1}{\rho_0 (x)} \frac{d\rho (r)}{dr}\right)_{r=x},
\label{wfn}
\end{equation}
with $\int_0^{\infty} dx \vert F(x) \vert^2 =1$,
which is the basic quantity that relates NM parameters in $x-$space and finite nuclei in $r-$space. The $r-$ of finite nuclei and $x-$space of infinite nuclear matter are matched together in the LDA formalism with the superposition of the total density of the nucleus  and an infinite number of $Fluctons$ in the  CDFM approach. 
The incompressibility $K^{A}$  Eq. ($\ref{k0}$) are the surface weighted average of the corresponding NM quantity in the LDA limit for finite nuclei. To estimate these surface properties, the densities are obtained self-consistently from E-RMF and folded with the nuclear matter parameters using the CDFM.

 The incompressibility $K^A$ is related with a constrained approach \cite{Bohi1979,Maru1989,Boer1991,Stoi1994,Stoi1994a}. Since, the $K^A$ is a measure of the breathing mode oscillation, the constrained energy $E(\eta)$ is expanded in a harmonic approximation as
\begin{eqnarray}
E(\eta)& = &E(0)+\frac{\partial E(\eta)}{\partial \eta}\big|_{\eta =0}
+\frac{\partial^2{E(\eta)}}{\partial{\eta}^2}|_{\eta =0}.
\end {eqnarray}
The second-order derivative in the expansion is related to the constrained incompressibility ${K}^A$ for a finite nucleus with mass number $A$ as
$K^{A} = \frac{1}{A} {R_0}^{2} \frac{\partial^2{E \eta}}
{\partial{R_\eta}}$,
%
and the monopole excitation energy ${E_M}$ is written as
${{E_M}={\sqrt{\frac{ {K^A}}{B^m}}}}$,  
where $B_m$ is the mass parameter \cite{Patr2002, Patr2001}. 
The centroid energy of the ISGMR is connected with the incompressibility $K^A$ by considering the distribution of nucleons as a radial function, as \cite{Brue1970,Blai1976,Pate2012,Stri1982}:

\begin{eqnarray}
{{E_M}={\hbar}{\sqrt{\frac{ {K^A}}{m<r_{rms}^2>}}}}, 
\end{eqnarray}
where $<r_{rms}^2>$ is the mean square radius of the nucleus in the ground state and $m$ is the mass of the nucleon. 
\begin{figure}
\centering
\includegraphics[width=0.8\columnwidth]{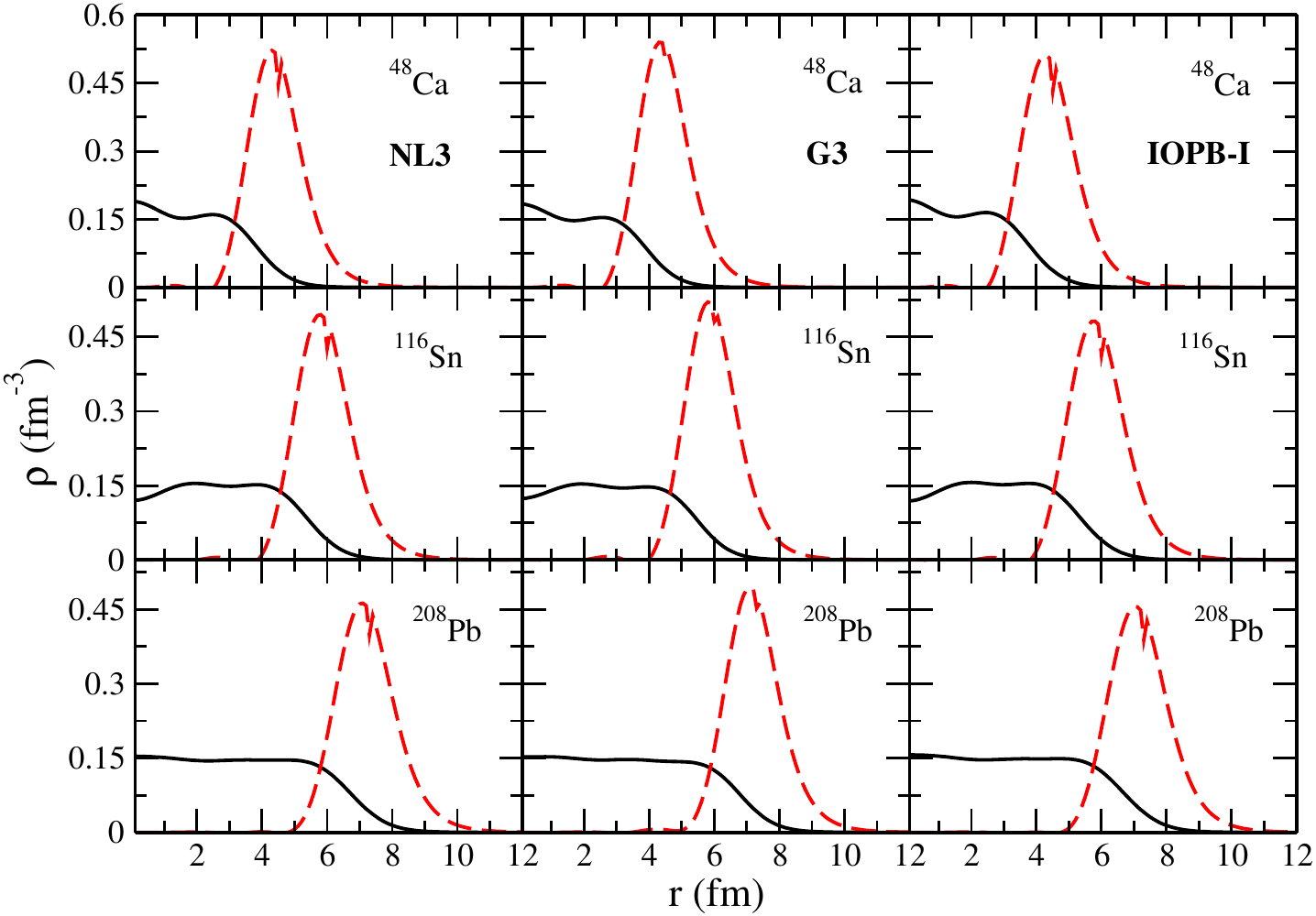}
\includegraphics[width=0.8\columnwidth]{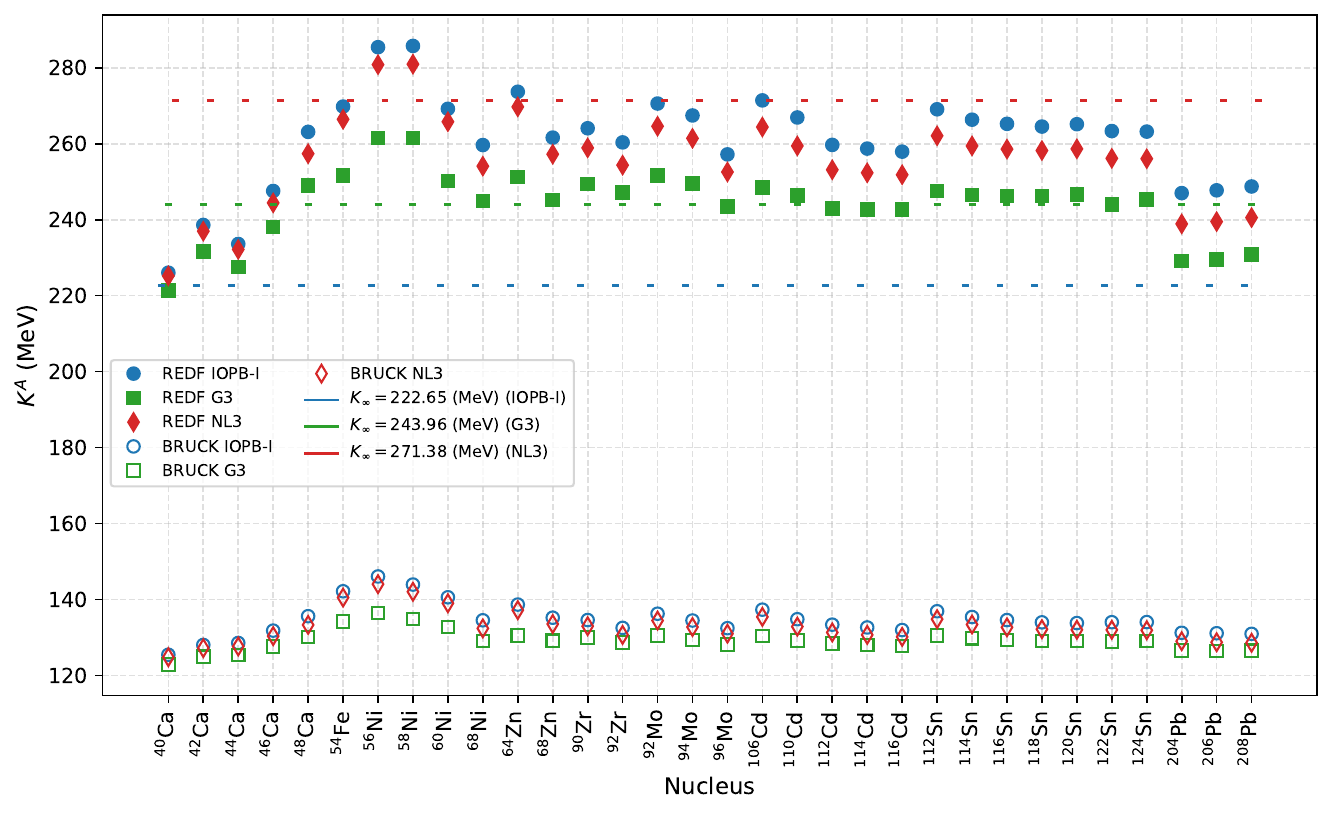}
\caption{(Upper Panel) The nucleon density distributions for $^{48}$Ca, $^{116}$Sn and $^{208}$Pb and their respective weight functions $|F(x)|^2$. (Lower Panel) The finite nucleus incompressibility for some selected nuclei with NL3, G3 and IOPB-I parameter sets.}
\label{fig1}
\end{figure}
Now, we present the results of our calculations. First, we calculate the nuclear matter energy density functional in momentum space. This functional then translates to the coordinate space in a local density approximation. Here we get an analytical expression of energy density in the position space. From this expression, we get various expressions of nuclear matter observables, such as incompressibility $K_{\infty}$. Before going to the $K^A$ and $E_M$ calculations, we calculate the density of the finite nuclei. The results of our investigations are given in Table \ref{tab1} and Figure \ref{fig1}. All the calculations are done with the recently developed extended relativistic mean field formalism with the use of the most successful NL3, IOPB-I, and G3 parameter sets \cite{Lala1997,kuma17,kuma18} in the relativistic Hartree approximation. The calculation is also extended to the Br\"uckner's energy density functional with the E-RMF densities obtained from the NL3, G3, and IOPB-I forces.   
 
\begin{table}
\caption{The iso-scalar giant monopole resonance $E_M$ obtained from the calculated incompressibility for G3, IOPB-I, and NL3 parameter sets. The results are obtained from the Br\"uckner (BRUCK) and relativistic energy density mean field formalism (REDF) approaches. The root mean square radius $r_{rms}$ is in fm and the energy is in MeV. The experimental monopole energy Expt. ($E_M$) is given in the last column \cite{biswal14,Gaid2025}.}
\label{tab1}
\renewcommand{\tabcolsep}{0.10cm}
\renewcommand{\arraystretch}{0.8}
\centering
\begin{tabular}{|ll|lll|ll|ll|ll|ll|}
\hline
\multicolumn{2}{|l|}{Nucleus} & \multicolumn{3}{c|}{$r_{rms}$} & \multicolumn{2}{c|}{G3 ($E_M$)} & \multicolumn{2}{c|}{IOPB-I ($E_M$)} & \multicolumn{2}{c|}{NL3 ($E_M$)} & \multicolumn{2}{c|}{Expt. ($E_M$)} \\
\hline
& & G3 & IOPB-I & NL3 & REDF & BRUCK & REDF & BRUCK & REDF & BRUCK & Value & Error \\
\hline
\multicolumn{2}{|l|}{$^{40}$Ca}  & 3.376 & 3.346 & 3.356 & 28.38 & 21.14 & 28.94 & 21.56 & 28.79 & 21.44 & \multicolumn{2}{l|}{19.18 $\pm$ 0.37} \\
\multicolumn{2}{|l|}{$^{42}$Ca}  & 3.423 & 3.395 & 3.404 & 28.63 & 21.03 & 29.3  & 21.46 & 29.12 & 21.33 & \multicolumn{2}{l|}{19.7 $\pm$ 0.1}   \\
\multicolumn{2}{|l|}{$^{44}$Ca}  & 3.467 & 3.438 & 3.447 & 28.02 & 20.8  & 28.63 & 21.24 & 28.47 & 21.11 & \multicolumn{2}{l|}{19.49 $\pm$ 0.34} \\            
\multicolumn{2}{|l|}{$^{48}$Ca}  & 3.542 & 3.5   & 3.519 & 28.69 & 20.74 & 29.84 & 21.42 & 29.36 & 21.13 & \multicolumn{2}{l|}{19.88 $\pm$ 0.16} \\
\hline
\multicolumn{2}{|l|}{$^{54}$Fe}  & 3.64  & 3.576 & 3.587 & 28.06 & 20.5  & 29.58 & 21.47 & 29.3  & 21.28 & \multicolumn{2}{l|}{19.66 $\pm$ 0.37} \\
\hline
\multicolumn{2}{|l|}{$^{56}$Ni}  & 3.671 & 3.597 & 3.61  & 28.37 & 20.49 & 30.25 & 21.64 & 29.89 & 21.41 & \multicolumn{2}{l|}{19.1$\pm$0.5}     \\
\multicolumn{2}{|l|}{$^{58}$Ni}  & 3.725 & 3.663 & 3.676 & 27.96 & 20.07 & 29.72 & 21.09 & 29.36 & 20.88 & \multicolumn{2}{l|}{18.43$\pm$0.15}   \\
\multicolumn{2}{|l|}{$^{60}$Ni}  & 3.775 & 3.723 & 3.736 & 26.98 & 19.65 & 28.38 & 20.51 & 28.1  & 20.32 & \multicolumn{2}{l|}{17.62$\pm$0.15}   \\
\multicolumn{2}{|l|}{$^{68}$Ni}  & 3.952 & 3.924 & 3.944 & 25.5  & 18.51 & 26.44 & 19.03 & 26.03 & 18.79 & \multicolumn{2}{l|}{21.1$\pm$1.9}     \\
\hline
\multicolumn{2}{|l|}{$^{64}$Zn}  & 3.871 & 3.835 & 3.845 & 26.37 & 19.01 & 27.78 & 19.77 & 27.5  & 19.62 & \multicolumn{2}{l|}{18.88$\pm$0.79}   \\
\multicolumn{2}{|l|}{$^{68}$Zn}  & 3.95  & 3.92  & 3.933 & 25.53 & 18.53 & 26.57 & 19.1  & 26.26 & 18.92 & \multicolumn{2}{l|}{16.6$\pm$0.17}    \\
\hline
\multicolumn{2}{|l|}{$^{90}$Zr}  & 4.281 & 4.248 & 4.261 & 23.76 & 17.15 & 24.63 & 17.59 & 24.32 & 17.43 & \multicolumn{2}{l|}{16.9$\pm$0.1}     \\
\multicolumn{2}{|l|}{$^{92}$Zr}  & 4.325 & 4.298 & 4.312 & 23.41 & 16.89 & 24.18 & 17.25 & 23.82 & 17.07 & \multicolumn{2}{l|}{16.5$\pm$0.1}     \\
\hline
\multicolumn{2}{|l|}{$^{92}$Mo}  & 4.308 & 4.271 & 4.282 & 23.72 & 17.08 & 24.8  & 17.6  & 24.46 & 17.44 & \multicolumn{2}{l|}{16.6$\pm$0.1}     \\
\multicolumn{2}{|l|}{$^{94}$Mo}  & 4.349 & 4.318 & 4.33  & 23.39 & 16.84 & 24.39 & 17.29 & 24.05 & 17.13 & \multicolumn{2}{l|}{16.4$\pm$0.2}     \\
\multicolumn{2}{|l|}{$^{96}$Mo}  & 4.388 & 4.362 & 4.375 & 22.9  & 16.61 & 23.68 & 16.99 & 23.39 & 16.85 & \multicolumn{2}{l|}{16.3$\pm$0.2}     \\
\hline
\multicolumn{2}{|l|}{$^{106}$Cd} & 4.522 & 4.49  & 4.499 & 22.45 & 16.26 & 23.63 & 16.81 & 23.27 & 16.66 & \multicolumn{2}{l|}{16.27 $\pm$ 0.09} \\
\multicolumn{2}{|l|}{$^{110}$Cd} & 4.582 & 4.555 & 4.568 & 22.06 & 15.98 & 23.1  & 16.41 & 22.71 & 16.25 & \multicolumn{2}{l|}{15.94 $\pm$ 0.07} \\
\multicolumn{2}{|l|}{$^{112}$Cd} & 4.61  & 4.585 & 4.6   & 21.77 & 15.83 & 22.63 & 16.22 & 22.27 & 16.04 & \multicolumn{2}{l|}{15.80 $\pm$ 0.05} \\
\multicolumn{2}{|l|}{$^{114}$Cd} & 4.641 & 4.621 & 4.637 & 21.62 & 15.7  & 22.42 & 16.05 & 22.06 & 15.88 & \multicolumn{2}{l|}{15.61 $\pm$ 0.08} \\
\multicolumn{2}{|l|}{$^{116}$Cd} & 4.671 & 4.655 & 4.672 & 21.48 & 15.58 & 22.22 & 15.89 & 21.87 & 15.72 & \multicolumn{2}{l|}{15.44 $\pm$ 0.06} \\
\hline
\multicolumn{2}{|l|}{$^{112}$Sn} & 4.6   & 4.567 & 4.579 & 22.03 & 15.99 & 23.13 & 16.5  & 22.77 & 16.33 & \multicolumn{2}{l|}{16.2 $\pm$ 0.1}   \\
\multicolumn{2}{|l|}{$^{114}$Sn} & 4.627 & 4.596 & 4.609 & 21.85 & 15.85 & 22.87 & 16.3  & 22.5  & 16.14 & \multicolumn{2}{l|}{16.1 $\pm$ 0.1}   \\
\multicolumn{2}{|l|}{$^{116}$Sn} & 4.656 & 4.63  & 4.644 & 21.7  & 15.73 & 22.65 & 16.13 & 22.3  & 15.97 & \multicolumn{2}{l|}{15.8 $\pm$ 0.1}   \\
\multicolumn{2}{|l|}{$^{118}$Sn} & 4.684 & 4.662 & 4.677 & 21.58 & 15.62 & 22.47 & 15.99 & 22.12 & 15.83 & \multicolumn{2}{l|}{15.8 $\pm$ 0.1}   \\
\multicolumn{2}{|l|}{$^{120}$Sn} & 4.714 & 4.693 & 4.709 & 21.45 & 15.52 & 22.34 & 15.87 & 21.99 & 15.71 & \multicolumn{2}{l|}{15.7 $\pm$ 0.1}   \\
\multicolumn{2}{|l|}{$^{122}$Sn} & 4.74  & 4.72  & 4.739 & 21.22 & 15.43 & 22.14 & 15.79 & 21.75 & 15.61 & \multicolumn{2}{l|}{15.4 $\pm$ 0.1}   \\
\multicolumn{2}{|l|}{$^{124}$Sn} & 4.765 & 4.745 & 4.766 & 21.16 & 15.35 & 22.02 & 15.71 & 21.62 & 15.52 & \multicolumn{2}{l|}{15.3 $\pm$ 0.1}   \\
\hline
\multicolumn{2}{|l|}{$^{204}$Pb} & 5.589 & 5.572 & 5.593 & 17.44 & 12.96 & 18.16 & 13.24 & 17.8  & 13.08 & 13.98                      &                      \\
\multicolumn{2}{|l|}{$^{206}$Pb} & 5.605 & 5.588 & 5.612 & 17.41 & 12.92 & 18.14 & 13.2  & 17.76 & 13.02 & 13.94                      &                      \\
\multicolumn{2}{|l|}{$^{208}$Pb} & 5.624 & 5.609 & 5.636 & 17.39 & 12.88 & 18.11 & 13.14 & 17.72 & 12.96 & \multicolumn{2}{l|}{13.96 $\pm$ 0.2}\\
\hline
\label{tab1}
\end{tabular}
\end{table}
The density distributions $\rho$ for some of the specific nuclei, such as $^{48}$Ca, $^{116}$Sn, and $^{208}$Pb, as a function of radius, are shown in Figure \ref{fig1} for the NL3 parameter set. Similar results are shown for the same isotopes with IOPB-I and G3 forces, also displayed in the left panel of the figure. The corresponding weight functions are also given in the same figure. The shell structure of the nuclei is clearly visible in all cases. A more careful inspection reveals that the central density of the Ca isotope is more than the average nuclear matter density, showing the presence of more nucleons at the center. On the other hand, the weight function is shown on the right side of the figure. The shape of the weight function $|F(x)|^2$ resembles a Gaussian-type distribution centering towards the tail region of the density. That means, the contribution of the weight function is maximum near the surface of the nucleus.
From the calculated density, we estimate the weight function $|F(x)|^2$ and show it in the same figure of density. The peak of the weight function mostly overlaps with the tail part of the density. Thus, it is clear that the major contribution of the observable comes from the surface region of the nucleus. Similar trends are observed with the IOPB-I and G3 parameter sets, indicating that the density distribution and weight function of the nuclei exhibit force-independent characteristics. The density distribution of the nucleons, along with the weight function for other considered nuclei, is also calculated.
 
The incompressibility for a finite nucleus $K^A$ is calculated using equation (3). From the known expression of $K_{\infty}$ in an $x-$space of infinite nuclear matter, we compute the $K^A$ for a finite nucleus using the coherent density fluctuation model (CDFM) by folding through the weight function $|F(x)|^2$. The computed values of $K^A$ are displayed in Table \ref{tab1} for G3, IOPB-I, and NL3 forces. The  $K^A$ are evaluated with the relativistic energy density functional as well as with the local density approximation of Br\"uckner approach with the same parameter sets. To our surprise, although the incompressibility at nuclear matter saturation $K_{\infty}$ is maximum for the NL3 set, we get the minimum $K^A$ for finite nuclear systems, which can be seen in Table \ref{tab1} and the right panel of the figure. Again, the REDF predicts an overestimation of $K^A$ against the Br\"uckner's approach. For example, $^{40}$Ca has an incompressibility of 221.33 MeV with REDF-G3, whereas it is 122.88 MeV with BRUCK-G3 (See Table \ref{tab1}). Similar results can also be noted for all other isotopes. To get more insight on it, we noted down the incompressibility at the nuclear matter saturation, i.e., the values of $K_{\infty}$ for G3, IOPB-I, and NL3 are found to be 243.96, 222.65, and 271.38 MeV, respectively. Interestingly, the finite nuclear incompressibility $K^A$ is just the reverse, i.e., IOPB-I gives the highest $K^A$ and NL3 predicts the lowest, which can be seen in Table \ref{tab1}.

The monopole excitation energy obtained from the CDFM approach is tabulated in Table \ref{tab1} for some experimentally known nuclei. The computed values for relativistic energy density functional with G3, IOPB-I, and NL3 sets are compared with the Br\"uckner's functionals. In almost all cases, the REDF overestimates the BRUCK results, irrespective of the force parameters. The ISGMR of the NL3 set is found to be the lowest, and that computed with the G3 is the highest in the series. These ISGMRs are evaluated using the formula of equation (9). In the expression, mostly the iso-scalar giant monopole resonances depend on the incompressibility of the finite nucleus along with it's size. Although, the radius of the nucleus is a fixed quantity,
it's incompressibility $K^A$ depends extensively on the hardness or softness of the nucleus.  Thus, it is interesting to note that, being the highest value of $K_{\infty}$ for the NL3 force, similar to the case of incompressibility, we find a minimum  ISGMR for this set. On the other hand, IOPB-I predicts the highest excitation energy for almost all the isotopes. This means, although the $K_{\infty}$ is the lowest for IOPB-I force, it is most compressible for finite nuclear system, which gives a comparatively larger excitation energy. 

This trend of ISGMR or $K^A$ is quite different than the normal findings. In Refs. \cite{Patr2001,Patr2002}, it is shown that higher the $K_{\infty}$, larger is the giant monopole excitation energy for finite nuclei. However, in the present investigation, we find opposite behavior, we get a minimum $E_M$, for the NL3 set, which has a larger nuclear matter incompressibility of 271 MeV. This reverse nature than the normal case may be due to the non-linearity in the vector meson interactions. This may be the self- and cross-couplings of the vector-mesons interactions. The normal trend (increasing of $K^A$ with $K_{\infty}$)  is valid as long as there is no nonlinearity in the potential of the vector components of the finite nucleus \cite{Patr2001,Patr2002}. However, this scenario changes
when some vector component is added to the nuclear potential. In the cases of G3 and IOPB-I, the non-linear self- and cross-couplings of the vector mesons are included in the interaction Lagrangian, as a result, we get a vector non-linear part in the nuclear potential, which could be the main driving effect for the modification of the EoS as well as the incompressibilities of the nuclear matter at saturation and finite nuclei $K^A$. 


In summary, the present work investigates the finite nuclei incompressibility and the corresponding isoscalar giant monopole resonance energies using the CDFM approach with both relativistic and Br\"uckner energy density functionals. Contrary to the conventional expectation that higher nuclear matter incompressibility leads to larger $K^A$ and ISGMR energy, our results show an inverse trend that the NL3 set, although having the highest $K_{\infty}$, produces the lowest $K^A$ and ISGMR, while the IOPB-I force predicts the highest values of these quantities.
This anomaly highlights the critical role of nonlinear vector-meson interactions. Specifically, the self- and cross-couplings of vector mesons, present in G3 and IOPB-I parameter sets, introduce a non-linear vector potential that significantly affects the density distributions and stiffness of the nuclear matter in finite systems. These findings suggest that the traditional correlation between $K_{\infty}$ and the ISGMR energy may not hold in the presence of such non-linearities. Therefore, it is suggested that while interpreting isoscalar giant monopole resonance energies, one has to be careful to separate the effects that come from the finite size of the nucleus and those that arise from the specific nature of the nuclear interactions.


\end{document}